# Multi-core processors – An overview

Balaji Venu[1]

*1 Department of Electrical Engineering and Electronics, University of Liverpool, Liverpool, UK*

**Abstract**

Microprocessors have revolutionized the world we live in and continuous efforts are being made to manufacture not only faster chips but also smarter ones. A number of techniques such as data level parallelism, instruction level parallelism and hyper threading (Intel's HT) already exists which have dramatically improved the performance of microprocessor cores. [1, 2] This paper briefs on evolution of multi-core processors followed by introducing the technology and its advantages in today's world. The paper concludes by detailing on the challenges currently faced by multi-core processors and how the industry is trying to address these issues.

## 1. Introduction

The microprocessor industry continues to have great importance in the course of technological advancements ever since their coming to existence in 1970s [3]. The growing market and the demand for faster performance drove the industry to manufacture faster and smarter chips. One of the most classic and proven techniques to improve performance is to clock the chip at higher frequency which enables the processor to execute the programs in a much quicker time [4, 5] and the industry has been following this trend from 1983 – 2002 [6]. Additional techniques have also been devised to improve performance including parallel processing, data level parallelism and instruction level parallelism which have all proven to be very effective [1]. One such technique which improves significant performance boost is multi-core processors. Multi-core processors have been in existence since the past decade, but however have gained more importance off late due to technology limitations single-core processors are facing today [7] such as high throughput and long lasting battery life with high energy efficiency [6].

## 2. Evolution of Multi-core processor

Driven by a performance hungry market, microprocessors have always been designed keeping performance and cost in mind. Gordon Moore, founder of Intel Corporation predicted that the number of transistors on a chip will double once in every 18 months to meet this ever growing demand which is popularly known as Moore's Law in the semiconductor industry [5, 8, 9]. Advanced chip fabrication technology alongside with integrated circuit processing technology offers increasing integration density which has made it possible to integrate one billion transistors on a chip to improve performance [7, 10]. However, the performance increase by micro-architecture governed by Pollack's rule is roughly proportional to square root of increase in complexity [11]. This would mean that doubling the logic on a processor core would only improve the performance by 40%. With advanced chip fabrication techniques comes along another major bottleneck, power dissipation issue. Studies have shown that transistor leakage current increases as the chip size shrinks further and further which increases static power dissipation to large values as shown in Figure 1 [11, 12]. One alternate means of improving performance is to increase the frequency of operation which enables faster execution of programs [4, 5]. However the frequency is again limited to 4GHz currently as any increase beyond this frequency increases power dissipation again [5]. "Battery life and system cost constraints drive the design team to consider power over performance in such a scenario" [13]. Power consumption has increased to such high levels that traditional air-cooled microprocessor server boxes may require budgets for liquid-cooling or refrigeration hardware [13]. Designers eventually hit what is referred to as the power wall, the limit on the amount of power a microprocessor could dissipate [14].

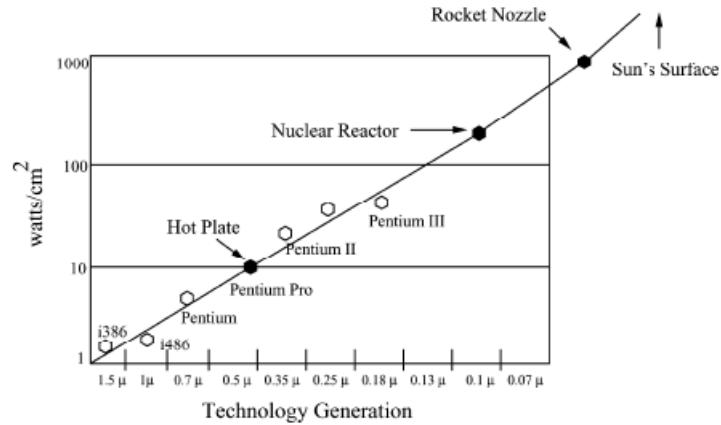

Figure 1: Power density rising [15]

Semiconductor industry once driven by performance being the major design objective, is today being driven by other important considerations such chip fabrication costs, fault tolerance, power efficiency and heat dissipation [9]. This led to the development of multi-core processors which have been effective in addressing these challenges.

### 3. Multi-core processors

"A Multi-core processor is typically a single processor which contains several cores on a chip" [7]. The cores are functional units made up of computation units and caches [7]. These multiple cores on a single chip combine to replicate the performance of a single faster processor. The individual cores on a multi-core processor don't necessarily run as fast as the highest performing single-core processors, but they improve overall performance by handling more tasks in parallel [16]. The performance boost can be seen by understanding the manner in which single core and multi-core processors execute programs. Single core processors running multiple programs would assign time slice to work on one program and then assign different time slices for the remaining programs. If one of the processes is taking longer time to complete then all the rest of the processes start lagging behind. [16] However, In the case of multi-core processors if you have multiple tasks that can be run in parallel at the same time, each of them will be executed by a separate core in parallel thus boosting the performance as shown in figure 2 [16].

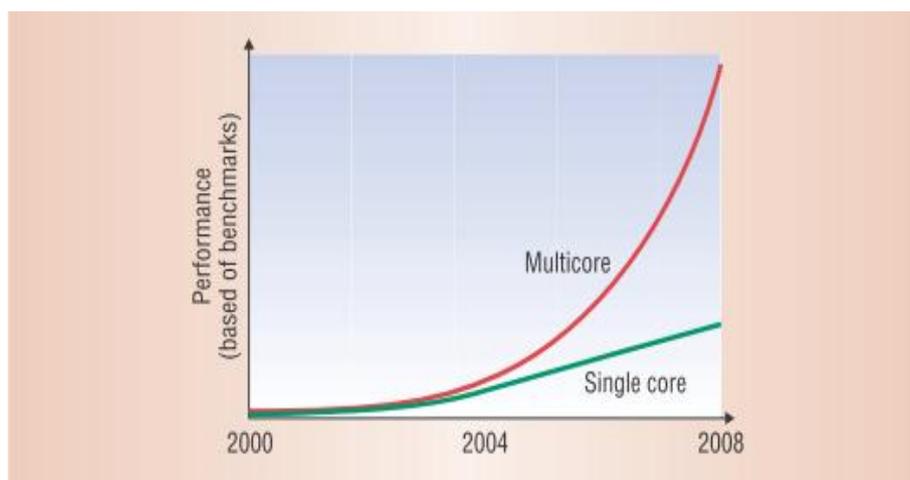

Figure 2. Multicore chips perform better – based on Intel tests using the SPECint2000 and SPECfp2000 benchmarks – than single-core processors. [16]

The multiple cores inside the chip are not clocked at a higher frequency, but instead their capability to execute programs in parallel is what ultimately contributes to the overall performance making them more energy efficient and low power cores as shown in the figure below [6]. Multi-core processors are generally designed partitioned so that the unused cores can be powered down or powered up as and when needed by the application contributing to overall power dissipation savings.

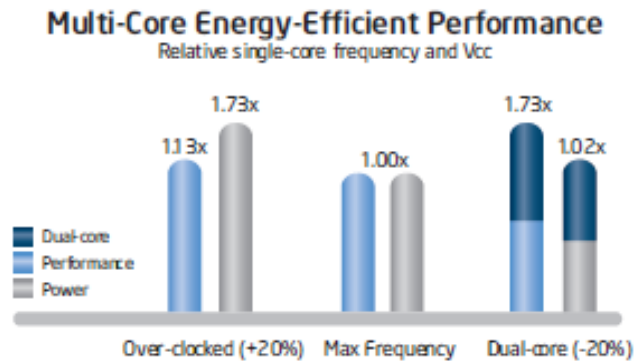

Figure 3: Dual core processor at 20% reduced clock frequency effectively delivers 73% more performance while approximately using the same power as a single-core processor at maximum frequency. [6]

Multi-core processors could be implemented in many ways based on the application requirement. It could be implemented either as a group of heterogeneous cores or as a group of homogenous cores or a combination of both. In homogeneous core architecture, all the cores in the CPU are identical [17] and they apply divide and conquer approach to improve the overall processor performance by breaking up a high computationally intensive application into less computationally intensive applications and execute them in parallel [5]. Other major benefits of using a homogenous multi-core processor are reduced design complexity, reusability, reduced verification effort and hence easier to meet time to market criterion [18]. On the other hand heterogeneous cores consist of dedicated application specific processor cores that would target the issue of running variety of applications to be executed on a computer [5]. An example could be a DSP core addressing multimedia applications that require heavy mathematical calculations, a complex core addressing computationally intensive application and a remedial core which addresses less computationally intensive applications [5].

Multi-core processors could also be implemented as a combination of both homogeneous and heterogeneous cores to improve performance taking advantages of both implementations. CELL multi-core processor from IBM follows this approach and contains a single general purpose microprocessor and eight similar area and power efficient accelerators targeting for specific applications has proven to be performance efficient [5].
"Another major multi-core benefit comes from individual applications optimized for multi-core processors. These applications when properly programmed, can split a task into multiple smaller tasks and run them in parallel [6]". Due to the multiple advantages that multi-core processors come along with, most of the processor manufactures started developing them. Intel announced that all its future processors will be multi-core when they realized that this technology can get past the power wall to improve performance [19, 20]. Other popular processor manufactures namely AMD, IBM and TENSILICA all have started developing multi-core processors. The number of cores in a processor is expected to increase and some even predict it to follow Moore's law [20]. TENSILICA's CEO has expressed his view on this technology stating that "System on a Chip will become a sea of processors. You will have ten to maybe a thousand processors on a chip" [21].

However it was observed that there is no throughput improvement while executing sequential programs/single threaded applications on multi-core chips due to under utilization of cores. Compilers are being developed which automatically parallelize applications so that multiple independent tasks can run simultaneously on different cores [10].

## 4. Major challenges faced by multi-core processors.

In spite of the many advantages that multi-core processors come with, there are a few major challenges the technology is facing. One main issue seen is with regard to software programs which run slower on multi-core processors when compared to single core processors. It has been correctly pointed out that "Applications on multi-core systems don't get faster automatically as cores are increased" [4]. Programmers must write applications that exploit the increasing number of processors in a multi-core environment without stretching the time needed to develop software [14]. Majority of applications used today were written to run on only a single processor, failing to use the capability of multi-core processors [22]. Although software firms can develop software programs capable of utilizing the multi-core processor to the fullest, the grave challenge the industry faces is how to port legacy software programs developed years ago to multi-core aware software programs [22]. Redesigning programs although sounds possible, it's really not a technological decision in today's environment. It's more of a business decision where in companies have to decide whether to go ahead redesigning software programs keeping in mind key parameters such as time to market, customer satisfaction and cost reduction [22].

The industry is addressing this problem by designing compilers which can port legacy single core software programs to 'multi-core aware' programs which will be capable of utilizing the power of multi-core processors. The compilers could perform "code reordering", where in compilers will generate code, reordering instructions such that instructions that can be executed in parallel are close to each other [10]. This would enable us to execute instructions in parallel improving performance. Also compilers are being developed to generate parallel threads or processes automatically for a given application so that these processes can be executed in parallel [10]. Intel released major updates for C++ and Fortran tools which aimed at programmers exploiting parallelism in multi-core processors. Also along side OpenMP (Open Multiprocessing), an application programming interface which supports multiprocessing programming in C, C++ and Fortran provides directives for efficient multithreaded codes [23]. It has however been correctly pointed out that "The throughput, energy efficiency and multitasking performance of multi-core processors will all be fully realized when application code is multi-core ready" [6].

Secondly, on-chip interconnects are becoming a critical bottle-neck in meeting performance of multi-core chips [24]. With increasing number of cores comes along the huge interconnect delays (wire delays) when data has to be moved across the multi-core chip from memories in particular [25]. The performance of the processor truly depends on how fast a CPU can fetch data rather than how fast it can operate on it to avoid data starvation scenario [26]. Buffering and smarter integration of memory and processors are a few classic techniques which have attempted to address this issue. Network on a chip (NoCs) are IPs (Intellectual Property) being developed and researched upon which are capable of routing data on a SoC in a much more efficient manner ensuring less interconnect delay [25].

Increased design complexity due to possible race conditions as the number of cores increase in a multi-core environment. "Multiple threads accessing shared data simultaneously may lead to a timing dependent error known as data race condition" [27]. In a multi-core environment data structure is open to access to all other cores when one core is updating it. In the event of a secondary core accessing data even before the first core finishes updating the memory, the secondary core faults in some manner. Race conditions are especially difficult to debug and cannot be detected by inspecting the code, because they occur randomly. Special hardware requirement implementing mutually exclusion techniques have to be implemented for avoiding race conditions [27].

Another important feature which impacts multi-core performance is the interaction between on chip components viz. cores, memory controllers and shared components viz. cache and memories [9] where bus contention and latency are the key areas of concern. Special crossbars or mesh techniques have been implemented on hardware to address this issue [9].

## 5. Conclusions

Power and frequency limitations observed on single core implementations have paved the gateway for multi-core technology and will be the trend in the industry moving forward. However the complete performance throughput can be realized only when the challenges multi-core processors facing today are fully addressed. A lot of technological breakthroughs are expected in this area of technology including a new multi-core programming language, software to port legacy software to "multi-core aware" software programs. Although it has been one of the most challenging technologies to adopt to, there is considerable amount of research going on in the field to utilize multi-core processors more efficiently.